\input{aipcheck}
\documentclass[final]{aipproc}
\layoutstyle{8x11single}
\begin{document}

\title{Composite quantum systems and environment-induced heating}

\author{Almut Beige}{
  address={The School of Physics and Astronomy, University of Leeds, Leeds LS2 9JT, United Kingdom}}

\author{Andreas Kurcz}{
  address={The School of Physics and Astronomy, University of Leeds, Leeds LS2 9JT, United Kingdom}}

\author{Adam Stokes}{
  address={The School of Physics and Astronomy, University of Leeds, Leeds LS2 9JT, United Kingdom}}

\begin{abstract}
In recent years, much attention has been paid to the development of techniques which transfer trapped particles to very low temperatures. Here we focus our attention on a heating mechanism which contributes to the finite temperature limit in laser sideband cooling experiments with trapped ions. It is emphasized that similar heating processes might be present in a variety of composite quantum systems whose components couple individually to different environments. For example, quantum optical heating effects might contribute significantly to the very high temperatures which occur during the collapse phase in sonoluminescence experiments. It might even be possible to design composite quantum systems, like atom-cavity systems, such that they continuously emit photons even in the absence of external driving.  
\end{abstract}

\maketitle

\section{Introduction}

Laser cooling allows or the transfer of single trapped ions to very low temperatures \cite{Wineland}. Its idea is relatively straightforward: A laser field constantly excites the ion from its ground state into an excited state. Since the laser Rabi frequency seen by the ion depends on its position, the laser establishes a coupling not only between different electronic states. Each transition is accompanied by a significant change of the vibrational state of the particle, especially, when the detuning of the laser is close to the phonon frequency $\nu$ of the trapped particle. In case of a red-detuned laser field, the excitation of the ion is in general accompanied by the annihilation of a phonon. If such a transition is followed by the spontaneous emission of a photon, the ion returns into its ground state without regaining its previously lost vibrational energy. A phonon is permanently lost which implies cooling.

Laser sideband cooling experiments indeed report cooling of single ions down to temperatures close to absolute zero \cite{cooling}. Fortunately, in ion trap experiments, there are only very few heating mechanisms which limit the final temperature of the cooling process. One of them is recoil, i.e.~the creation of phonon energy during the spontaneous emission of a photon. One way to eliminate the effect of recoil, is to use cavities where phonon energy is removed from the system via the leakage of photons through the cavity mirrors \cite{Tony}. In cavity-based laser cooling, the finite temperature limit is only due to the counter-rotating terms in the cavity-phonon interaction Hamiltonian which result in the simultaneous and in general highly detuned excitation of electronic {\em and} vibrational states. Without these terms, we predict cooling of trapped particles down to zero phonon numbers \cite{NJP}.

Although being an almost negligible effect in ion trap experiments, this paper focusses its attention on the above mentioned quantum optical heating mechanism. We attribute it to the counter-rotating terms in the interaction Hamiltonian of composite quantum system with the ability to emit photons. It is emphasized that similar quantum heating effects might be present in a variety of composite quantum systems whose components couple individually to different environments. They might contribute to the very high temperatures which occur during the collapse phase in sonoluminescence experiments \cite{sono}. It might even be possible to design composite quantum systems, like atom-cavity systems, such that they continuously emit photons even in the absence of external driving \cite{ec1,ec2}.   

Although the continuous emission of photons without an apparent external source is expected to occur only at a relatively low rate and does not violate the basic laws of thermodynamics  \cite{ec3}, its prediction might be an indication that we need to revisit our current approaches to the modeling of quantum optical systems with spontaneous emission beyond the rotating wave and certain Markovian approximations. For example, an environment might refuse to pump energy into an open quantum system. In this case, master equations for the system have to be derived using a so-called {\em lazy detector model}. The predictions of this model can be different from more conventional derivations of master equations \cite{Adam}. This is an interesting observation, since due to fast experimental progress \cite{Mooji}, it becomes increasingly important to model quantum optical systems more and more accurately. 

\section{Energy conservation in atomic processes}

Already in 1936, Dirac pointed out that the question of energy conservation in atomic processes is not a trivial question \cite{Dirac}. For example, in a {\em closed} quantum system with a Hermitian Hamiltonian $H$ which commutes with itself, the expectation value for the total energy of the system is a conserved quantity. However, this argument does not apply to {\em open} quantum systems, like quantum optical system with the capability to emit photons. Consider for example a single trapped ion with ground state $|0 \rangle$ and excited state $|1 \rangle$. Suppose the ion is initially in the state $|\psi \rangle = \alpha \, |0 \rangle + \beta \, |1 \rangle$ with non-zero coefficients $\alpha$ and $\beta$. Then there is then a finite probability, namely $|\beta |^2$, for the spontaneous emission of a photon. If such an emission occurs, a quanta of energy $\hbar \omega_0$ is released. With probability $|\alpha |^2$, no photon is emitted and the atomic state becomes $|0 \rangle $ without releasing any energy. Energy conservation applies only averaged over an ensemble of atoms. Over a short time interval, the atom-detector system is capable of borrowing energy from the vacuum.

In his book on the quantum vacuum \cite{Milonni}, Milonni calculates the acceleration for a linear dipole oscillator inside the vacuum. Starting from the minimal coupling Hamiltonian for the interaction between the particle and the surrounding free radiation field, he calculates the time evolution of the position of the particle. Although there is no external driving, he predicts an acceleration of the particle. Its origin too is the interaction between the electric dipole and the vacuum fields. We know that Milonni's predictions cannot be verified easily experimentally. Here we therefore discuss a closely related effects in composite quantum systems which might be already or might become soon accessible experimentally.

\begin{figure}[b]
  \includegraphics[height=.1\textheight]{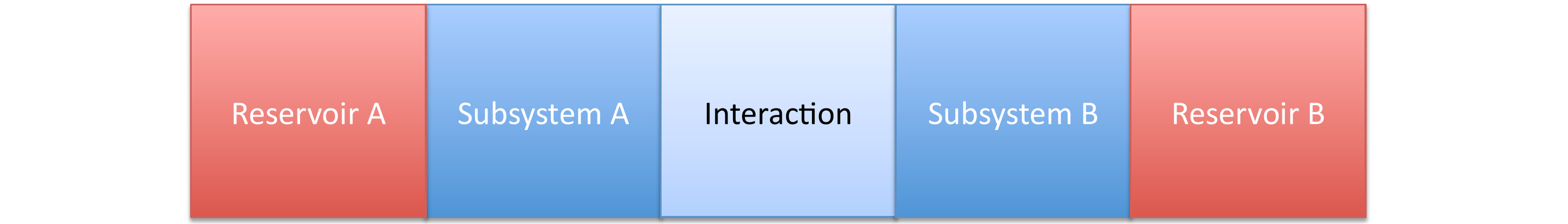} 
  \caption{Schematic view of a composite quantum system whose two components $A$ and $B$ interact individually with different environments. In the following we assume that the Hamiltonian for the interaction between both system components contains counter-rotating terms such that the ground state of the system becomes entangled.} \label{fig1}
\end{figure}

To do so, we now have a closer look at the composite quantum system illustrated in Fig.~\ref{fig1} whose two components interact with two different reservoirs. Without the interaction with these reservoirs, the Hamiltonian of the composite quantum system can be written as 
\begin{eqnarray} \label{Hsys}
H_{\rm sys} &=& H_A + H_B + H_{\rm int} \, .
\end{eqnarray}
If the interaction Hamiltonian $H_{\rm int}$ does not contain any counter-rotating terms, then the ground state $|\lambda_0 \rangle_{AB}$ of the composite quantum system is simply given by $|\lambda_0 \rangle_{AB} = |\lambda_0 \rangle_A \otimes |\lambda_0 \rangle_B$, where $|\lambda_0 \rangle_X$ denotes the ground state of subsystem $X$, i.e.~the ground state of the Hamiltonian $H_X$ in Eq.~(\ref{Hsys}). However, this no longer applies when the system Hamiltonian $H_{\rm int}$ contains counter-rotating terms. Then we have
\begin{eqnarray} \label{not}
|\lambda_0 \rangle_{AB} &\neq & |\lambda_0 \rangle_A \otimes |\lambda_0 \rangle_B \, .
\end{eqnarray}
The ground state of a composite quantum system with interacting system components $A$ and $B$ is in general {\em entangled}.  Suppose now that $A$ and $B$ interact individually with different environments,  as illustrated in
Fig.~\ref{fig1}. Moreover, we assume in the following that the effect of the reservoirs $A$ and $B$ constitutes a measurement of the energy of the respective subsystem. This assumption is often used when deriving master equations for quantum optical systems, where the emission of a photon or its absence indicates whether its source is in an excited state or not \cite{Hegerfeldt93,Dalibard,Carmichael}. After the detection of no photon in reservoir $A$ {\em and} in reservoir $B$, the state of the quantum system is in this case to a very good approximation given by $ |\lambda_0 \rangle_A \otimes |\lambda_0 \rangle_B$. A comparison with Eq.~(\ref{not}) shows that this state is not the ground state of the system Hamiltonian $H_{\rm sys}$ in Eq.~(\ref{Hsys}). Its energy expectation value $_B \langle \lambda_0| \otimes  _A \langle \lambda_0 | H_{\rm sys} |\lambda_0 \rangle_A \otimes |\lambda_0 \rangle_B $ is in fact higher than the
energy expectation value $_{AB} \langle \lambda_0 | H_{\rm sys} |\lambda_0 \rangle_{AB} $ of the ground state $|\lambda_0 \rangle_{AB}$ of the system. It therefore evolves in time, thereby resulting in the population of the energy eigenstates $|\lambda_i \rangle_A$ and $|\lambda_i \rangle_B$ of the two subsystems $A$ and $B$ with $i>0$. As a result, there is now a non-zero probability for the emission of a photon due to these subsystems being in excited energy
eigenstates. It might look as if energy appears from nowhere! 

The above described process, where a composite quantum system reaches a state with higher energy than its ground state energy by jumping over a separating barrier is only possible with the help of excitations from the vacuum. The environment pumps energy into the system when performing measurements on the different components of the composite quantum system. The reason for this is that these measurements are accompanied by the loss of correlations, i.e.~by the loss of entanglement, when preparing the system components $A$ and $B$ in a de-correlated state. Performing de-correlating measurements requires energy as it has already been pointed out earlier by Schulman and Gaveau \cite{Schulman} thereby fulfilling the laws of themodynamics \cite{Ford,Ford2}. Since the total energy of quantum system and environment is distributed into many degrees of freedom one requires that the total balance of energy and entropy remains constant in time, thereby keeping the free energy of the open system constant. 

\section{Concrete examples}

We now have a closer look at concrete examples of composite quantum systems with environment-induced heating. All three examples considered in have subsystems $A$ and $B$ which couple to separate environments such that the detection of a photon corresponds to a measurement of the energy of one subsystem but not of both. 

\subsubsection{Laser-sideband cooling of trapped ions}

Laser sideband cooling requires a red-detuned laser field whose detuning ideally equals the frequency of the vibrational mode of the trapped particle \cite{Wineland}. When the laser excites the ion from its electronic ground state into an excited state, the vibrational energy reduces in general by one phonon. Afterwards, the ion returns most likely into its ground state via the spontaneous emission of a photon, i.e.~without regaining the lost energy in its vibrational mode. The phonon is permanently lost which implies cooling.

When choosing an appropriate interaction picture, the system Hamiltonian of the laser-driven trapped ion becomes time independent and has exactly the same form as the Hamiltonian $H_{\rm sys}$ in Eq.~(\ref{Hsys}). Subsystem $A$ contains the electronic states $|0 \rangle$ and $|1 \rangle$ of the ion, while subsystem $B$ contains its quantised vibrational states -- the phonons. The interaction Hamiltonian $H_{\rm int}$ includes terms which excite the ion while annihilating a phonons and vice versa. But it also contains a counter-rotating term which excites the ions while creating a phonon. Without this term, the ground state of the ion-phonon system would be the $|0,0 \rangle$ state with no phonons and no population in excited electronic states. This means, even after the spontaneous emission into $|0,0 \rangle$, the system evolves in time. Due to the energy-concentrating mechanism described in the previous subsection the stationary state temperature of the ion remains slightly greater than zero, even in the absence of recoil heating.

\subsubsection{Sonoluminescence}

Sonoluminescence is the intriguing phenomenon of strong light flashes from tiny bubbles in a liquid \cite{Brenner}. The bubbles are driven by an ultrasonic wave and need to be filled with noble gas atoms. Approximating the emitted light by blackbody radiation indicates very high temperatures. Although sonoluminescence has been studied extensively, the origin of the sudden energy concentration within the bubble collapse phase remains controversial \cite{Flannigan}. 

Above we have seen that the detection of photons from a composite quantum system constitutes an environment-induced heating mechanism. In Ref.~\cite{sono} it has therefore been speculated that a similar heating mechanism contributes substantially to the heating in sonoluminescence experiments. During rapid bubble deformations the collapse phase, a weak but highly inhomogeneous electric field as might occur. The Hamiltonian which describes the time evolution of the nobel gas atoms on the relevant short time scales is therefore of the same form as the system Hamiltonian $H_{\rm sys}$ in Eq.~(\ref{Hsys}). Their situation becomes similar to the situation of trapped ions in laser cooling experiments. The energy eigenstates of the atom-phonon system are in general entangled. This implies that the energy in the bubble can increase rapidly during photon emission \cite{sono,Schulman}. 

\subsubsection{Energy concentration in atom-cavity systems}

A system which is intuitively more accessible and experimentally much easier to control than sonoluminescing bubbles experiments consists of atoms trapped inside an optical cavity. Recently, it has been suggested to study the possibility of environment-induced heating in this system \cite{ec1,ec2,ec3}. When considering atom cavity-systems, we notice that their system Hamiltonian is of the same form as $H_{\rm sys}$ in Eq.~(\ref{Hsys}). Subsystem $A$ contains the atoms while subsystem $B$ contains the cavity field mode. Taking the counter-rotating terms in the atom-cavity interaction Hamiltonian into account we find that the ground state of the system is a state with a small amount of population in excited atomic and in excited cavity photon states. As long as the leakage of photons through the cavity mirror is caused by the excitation of cavity mode and excitation from the atomic states is caused only by the excitation of electronic states, we expect that continuous observations of the surrounding free radiation field pump energy into the setup. One might therefore expect a continuous leakage of photons through the cavity mirrors even in the absence of external driving.

\section{Conclusions}

This paper focussed its attention on a heating mechanism which contributes to the finite temperature limit in laser \cite{Wineland} and in cavity-based laser cooling \cite{NJP,Tony} with trapped ions. It is emphasized that similar heating processes might be present in a variety of composite quantum systems whose components couple individually to different environments. For example, quantum optical heating effects might contribute significantly to the very high temperatures which occur during the collapse phase in sonoluminescence experiments \cite{sono}. It might even be possible to design composite quantum systems, like atom-cavity systems, such that they continuously emit photons even in the absence of external driving. However, these effects might also be a sign that we need to refine our methods when modeling open quantum systems beyond the rotating wave and certain Markovian approximations \cite{Adam}. 

\begin{theacknowledgments}
A.B. acknowledge many interesting and stimulating discussions with A. Capopulo,  E. Del Giudice, and G. Vitiello. This work has been supported by the UK Research Council EPSRC.
\end{theacknowledgments}

\bibliographystyle{aipproc}  

\end{document}